\documentclass[sigconf]{acmart}
\AtBeginDocument{%
  \providecommand\BibTeX{{%
    \normalfont B\kern-0.5em{\scshape i\kern-0.25em b}\kern-0.8em\TeX}}}

\usepackage{array}
\usepackage{xcolor}
\usepackage{lineno}
\usepackage{hyperref,wasysym}
\usepackage[utf8]{inputenc}
\usepackage{array,longtable}
\usepackage{booktabs}
\usepackage{appendix}

\newcommand{\myparagraph}[1]{\vspace{1mm}\noindent\textbf{#1}}

\usepackage[T1]{fontenc}
\usepackage{subcaption}

\usepackage{algorithm}
\usepackage{algorithmic}
\usepackage{graphicx}

\usepackage[tikz]{bclogo}

\newcounter{mycounter}

\newenvironment{takeaway}{\stepcounter{mycounter} \begin{bclogo}[couleur=gray!10, arrondi=0.1,
logo= ,
noborder=true]{Insight \themycounter}}{\end{bclogo}}
\usepackage{fontawesome5}

\usepackage{version}
\excludeversion{note}

\newtoggle{showmarks}
\toggletrue{showmarks}  
\renewenvironment{comment}{\note\ignorespaces}{\endnote\ignorespacesafterend}

\newcommand{\CompanyX}{Microsoft}

\iftoggle{showmarks}{
  
  \newcommand\anjaly[1]{\textcolor{blue}{Anjaly: #1}}
   \newcommand\ayush[1]{\textcolor{red}{Ayush: #1}}

      \newcommand\Suman[1]{\textcolor{red}{[Suman: #1]}}
 
                  \newcommand\Jonathan[1]{\textcolor{red!55!blue}{Jonathan: #1}}
                  
    \newcommand\TODO[1]{\textcolor{blue!55!green}{TODO: #1}}
  
}{
 
\newcommand\anjaly[1]{\unskip}  
\newcommand\ayush[1]{\unskip}  
\newcommand\Chetan[1]{\unskip}  
\newcommand\Suman[1]{\unskip}  
\newcommand\Jonathan[1]{\unskip}  
\newcommand\TODO[1]{\unskip}

}

\makeatletter
\def\@copyrightspace{\relax}
\makeatother

\begin{document}
\title{Intelligent Monitoring Framework for Cloud Services: A Data-Driven Approach}


\author{Pooja Srinivas, Fiza Husain, Anjaly Parayil, Ayush Choure, Chetan Bansal, Saravan Rajmohan}
\email{{t-psrinivas, t-fizahusain, aparayil,aychoure, chetanb, saravan.rajmohan}@microsoft.com}
\affiliation{%
  \institution{Microsoft}
  \country{} 
}
\begin{comment}
\author{Lars Th{\o}rv{\"a}ld}
\affiliation{%
  \institution{The Th{\o}rv{\"a}ld Group}
  \streetaddress{1 Th{\o}rv{\"a}ld Circle}
  \city{Hekla}
  \country{Iceland}}
\email{larst@affiliation.org}

\author{Valerie B\'eranger}
\affiliation{%
  \institution{Inria Paris-Rocquencourt}
  \city{Rocquencourt}
  \country{France}
}

\author{Aparna Patel}
\affiliation{%
 \institution{Rajiv Gandhi University}
 \streetaddress{Rono-Hills}
 \city{Doimukh}
 \state{Arunachal Pradesh}
 \country{India}}

\author{Huifen Chan}
\affiliation{%
  \institution{Tsinghua University}
  \streetaddress{30 Shuangqing Rd}
  \city{Haidian Qu}
  \state{Beijing Shi}
  \country{China}}

\author{Charles Palmer}
\affiliation{%
  \institution{Palmer Research Laboratories}
  \streetaddress{8600 Datapoint Drive}
  \city{San Antonio}
  \state{Texas}
  \country{USA}
  \postcode{78229}}
\email{cpalmer@prl.com}

\author{John Smith}
\affiliation{%
  \institution{The Th{\o}rv{\"a}ld Group}
  \streetaddress{1 Th{\o}rv{\"a}ld Circle}
  \city{Hekla}
  \country{Iceland}}
\email{jsmith@affiliation.org}

\author{Julius P. Kumquat}
\affiliation{%
  \institution{The Kumquat Consortium}
  \city{New York}
  \country{USA}}
\email{jpkumquat@consortium.net}

\renewcommand{\shortauthors}{Trovato and Tobin, et al.}
    
\end{comment}

\begin{abstract}
Cloud service owners need to continuously monitor their services to ensure high availability and reliability. Gaps in monitoring can lead to delay in incident detection and significant negative customer impact. Current process of monitor creation is ad-hoc and reactive in nature. Developers create monitors using their tribal knowledge and, primarily, a trial and error based process. As a result, monitors often have incomplete coverage which leads to production issues, or, redundancy which results in noise and wasted effort.

In this work, we address this issue by proposing an intelligent monitoring framework that recommends monitors for cloud services based on their service 
properties. We start by mining the attributes of 30,000+ monitors from 791 production services at \CompanyX{} and derive a structured ontology for monitors. We focus on two crucial dimensions: what to monitor
(resources) and which metrics to monitor. We conduct an extensive empirical study and derive key insights on the major classes of monitors employed by cloud services at Microsoft, their associated dimensions, and the interrelationship between service properties and this ontology. 
Using these insights, we propose a deep learning based framework that recommends monitors based on the service properties.
Finally, we conduct a user study with engineers from \CompanyX{} which demonstrates the usefulness
of the proposed framework.
The proposed framework along with the ontology driven projections, succeeded in creating production quality recommendations for majority of resource classes. This was also validated by the users from the study who rated the framework's usefulness as 4.27 out of 5.
\end{abstract}

\begin{comment}

\ccsdesc[500]{Do Not Use This Code~Generate the Correct Terms for Your Paper}
\ccsdesc[300]{Do Not Use This Code~Generate the Correct Terms for Your Paper}
\ccsdesc{Do Not Use This Code~Generate the Correct Terms for Your Paper}
\ccsdesc[100]{Do Not Use This Code~Generate the Correct Terms for Your Paper}
    
\end{comment}



\maketitle

\section{Introduction}
At \CompanyX{}, we have hyper-scale cloud, with 5000+ services deployed across 60+ regions and used by hundreds of millions of users. Ensuring continuous availability of these services is crucial for maintaining customer satisfaction and business revenue \cite{surianarayanan2019cloud}. Despite significant efforts to ensure reliability, production incidents or failures are inevitable and can have a negative impact on customers, as well as requiring significant engineering resources and manual effort to mitigate. Hence, the early detection and mitigation of incidents is essential for minimizing the impact on customers and reducing the associated costs. To address this, service providers continuously monitor service health and proactively detect and mitigate incidents before they can impact customers. 

\noindent The current process for creating monitors is a trial and error based approach, in which service owners add monitors based on their understanding of service architecture, major dependencies, and service-level agreements. Additionally, existing monitors are modified and new monitors are added based on incidents in production. However, this approach is not very effective. Firstly, the process is reactive in nature, meaning service owners may miss critical monitors until an incident occurs. Secondly, often they create redundant monitors which results in noisy alert and wasted effort.

In recent years, several empirical studies have been conducted to characterize the challenges with monitoring and incident resolution
of cloud services \cite{gao2018empirical, ghosh2022fight, bhardwaj2021comprehensive, gunawi2014bugs, gunawi2016does, liu2019bugs, wang2022autothrottle, zhao2020understanding, chen2020towards}, as well as on Service Level Objectives and their practical challenges \cite{mogul2019nines,ding2019characterizing,ding2020characterizing}. However, the prior works assume that the exact monitors to be added are already known, and there is not much work on what to monitor within a service and which metrics to monitor for the performance, efficiency, and reliability of cloud services.

In this paper, we holistically study the monitor attributes from 791 production services from \CompanyX{}, with the aim of developing a systematic framework for monitor creation. 
We begin by 
focusing on two major dimensions of the monitors: resource classes (what to monitor within a service) and SLO classes (which metrics are most representative of the performance, efficiency, and reliability of a cloud service). We start by mining monitor data for NLP signals and derive a structured ontology for the monitors. Furthermore, we conduct an extensive empirical study using the derived ontology and correlate it with the properties of cloud services that employ these monitors. 
The insights from our empirical study serve as a precursor for developing a monitor recommendation framework based on the similarity among services. 
To summarize, we make the following contributions in this paper:
\begin{itemize}
    \item We present a comprehensive empirical analysis with an emphasis on deriving a structured ontology for the monitors within cloud services.
    This is accomplished by focusing on two critical dimensions: what to monitor and which metrics to monitor.
    \item We  developed an ontology for monitors that encompasses two primary dimensions: resource classes and service level objectives to be monitored. 
    \item Through an extensive empirical study of the monitor space using the proposed ontology, we derived valuable insights for a monitor recommendation framework, such as the structural correlation between ontology classes and the influence of service properties on the monitor ontology.
    \item We propose a novel monitor recommendation framework using prototypical learning, which learns relevant features and prototypes representative of each monitor class. 
    \item We have conducted a quantitative evaluation of the recommendations and surveyed engineers from \CompanyX{}, demonstrating the efficacy of our monitor recommendation framework.
\end{itemize}

\section{Background} \label{sec:background}
In this section, we formalize the context for the intelligent monitoring problem. We start with a set of definitions covering the basic concepts in the service monitoring space. We then describe \CompanyX{}'s context, in which this study takes place. Finally, we define the monitor recommendation problem and our formulation.
\subsection{Definitions}
In the context of a micro-service ecosystems, we define the following terms.

\noindent \textit{Resource} is defined as the underlying resource on which a critical issue can potentially happen. These can be internal to a service's environment, say CPU, Paging Cache, Stack Size, etc. or external dependencies, such as databases, cloud storage, etc. Resources will be henceforth denoted by $r$.

\noindent \textit{Functionality-Group} is an encapsulation of a set of individual disparate resources.  E.g. a specific VM is a \textit{functionality-group} which encapsulates multiple resources such as CPU, RAM, storage, etc. Functionality-groups will be henceforth denoted by $F$ with necessary subscripts. Every functionality-group $F$ expands into a set of resources, $r(F) = \{r_{i}\}$.

\noindent \textit{Metric} refers to a time-series object which gets generated as resources are used, and consequently, is always associated with a resource. A mathematical expression defining the members of this time-series is implicit here. An example of a commonly used metric is the CPU utilization on a machine hosting the service VM. 
Individual metrics will be denoted by the small letter $m$. For a given metric $m$, the corresponding resource will be succinctly represented as $r_{m}$.

\noindent \textit{Alerting Logic} are the anomaly detection rules which operate on metrics and act as triggers for alert generation. E.g. CPU utilization > 90\% for last 30 time steps would trigger an alert when CPU is likely to become a bottleneck. They also have a severity level associated with each of the rules. Alerting logic statements are \textit{always} associated with some metric. \footnote{There is another component of alerting logic, referred to as \textit{dimension} in certain infrastructure contexts. This field basically determines the \textit{scope} of aggregation of the metric. E.g. API latency measurements will typically be aggregated over clusters in specific regions. We have not explicitly mentioned this field (and others like it) for the sake of brevity and focus.} Individual alerting logic statements will be referred by the symbol $A$. The corresponding metric and resource are represented, respectively, as $m_{A}$ and $r_{A}$.

\noindent \textit{Monitor} is defined as a collection of triplets of (resource, metric, alerting logic). A monitor will \textit{fire} an alert of particular severity is an alerting logic (of same severity) gets activated. We will use the capital letter $M$ to denote an individual monitor. Its constituting triplets are thus $\{r_{i}, m_{i}, A_{i}\}_{1\leq i \leq k}$ where a given $M$ is a collection of $k$ individual alerting logic statements.

\noindent We use the symbol $S$ to refer to an individual \textit{micro-service}. For brevity, we overload these notation and use all these symbols as unimodal operators to go from one category to another (wherever applicable). E.g. $M(S)$ is the set of all monitors associated with $S$. Similarly, $m(M)$ is the set of all metrics associated with a monitor $M$.

\noindent \textit{Dependencies} of a service are defined using upstream and downstream services. Upstream services refer to the services that the given service relies on, while downstream services refer to the services that consume from the given service. Upstream dependencies of a service $S$ are denoted by the set $\mathcal{U}(S)$ and downstream dependencies are denoted by $\mathcal{D}(S)$. 

\subsection{\CompanyX{} Service Monitor Data}

We contextualize this study in \CompanyX{}'s monitor and incident data environment. In particular, we restrict ourselves to a specific vertical of MS products, which corresponds to reasonably sized set of (micro-)services each of which is independently provisioned, monitored, and maintained by it's owning team. The dataset we use consists of 791 unique micro-services with 30,920 unique monitors comprising of 7403 unique metrics leveraging specific sampling types. Every monitor has a unique alerting logic as well. The dataset also includes upstream and downstream dependencies for services. On an average, each service has 43 upstream dependencies and 20 downstream dependencies. 
Additionally, the dataset includes the physical and logical components of each service. Examples of physical components include cluster, VM, Datacenter, etc., while logical components include Role, Environment, Tenant, Stamp, etc. On average, each service from the dataset has 29 components.



A contextual artifact of the Microsoft monitor data is that the underlying resource name itself doesn't occur explicitly in the data. There is an operational justification for this. Oftentimes, the underlying resource is referred to by a variable name (e.g. a variable pointing to a captive database) which is subject to change over time. If the alerting logic and metric are tied to this specific variable name, then every code change around the variable itself will lead to cascade of changes in monitor data. Decoupling monitor structure with code helps these issues. What we therefore have is a \textit{functionality-group}, $F$, which acts as a container for a group of individual resources which are identified only by the metrics being emitted by time series measurements on them. While the functionality-group may contain many types of resources, only some of these might be critical (from run-time issues perspective) and will have observant metrics being recorded. We restrict our attention to only these resources which are already covered under some metric measurement. 

Consequently, a single monitor in the MS dataset is comprised of triplets $\{F_{i}, m_{i}, A_{i}\}_{1 \leq i \leq k}$ \footnote{As mentioned before, this is a slight simplification of the monitor data structure}. 


\subsection{The Monitor Recommendation Problem}

The general decision of what to monitor (and which SLO type to cover) is a complex question often requiring either deep expertise in both the engineering and product environment of the service, or unfortunate past experience of high severity breakdowns pointing to critical points in the system which require monitoring. We posit the following core problem as the central question in the monitor creation space. 

\noindent \textbf{Monitor Recommendation Problem} Given a service $S$, it's underlying resources $r(S)$ and dependencies $d(S)$, generate the set $M(S)$.

More succinctly we want to learn the association $M(S|r(S),d(S))$. However, this is an extremely hard problem to solve directly via simple association learning models for the many reasons, some of which are, 1) the underlying resources associated with a service aren't listed anywhere specific resources are too granular for ML models to learn transferable signals, 2) creating representations for services and resources which captures their functional nuances is a very hard problem by itself, and 3) predicting the optimal alerting logic subsumes the general anomaly detection problem and therefore cannot have a general solution. In fact in most scenarios, the threshold setting is more of an art than an exact science.
\begin{comment}
\begin{itemize}
    \item the underlying resources associated with a service aren't explicitly listed anywhere
    \item specific resources are too granular for ML models to learn transferable signals
    \item creating representations for services and resources which captures their functional nuances is a complex problem by itself
    \item predicting the optimal alerting logic subsumes the general anomaly detection problem and therefore cannot have a general solution. In fact in most scenarios, the threshold setting is more of an art than an exact science.
\end{itemize}
\end{comment}

In this paper, we simplify the problem to a more tractable version with the hope of incrementally reclaiming the complexity of the original problem back as the state of art improves. We make two core simplifications - (1) we forgo the alerting logic recommendation altogether, (2) instead of recommending the exact underlying resource, we introduce the concept of resource classes and SLO types. 

The \textit{resource classes} are defined as broad categories of resources that are \textit{typically} used by services. Common examples are CPU, RAM, Storage, APIs, etc. There is no universally accepted list of resource class which one can use to map every real resource to. We need to generate an appropriate list of these class labels which cover a big proportion of the resources in our dataset. Resource classes will henceforth be denoted by $C$. As before, we will overload as an operator to denote association. So, $C(r)$ denotes the resource class of the resource $r$, and $C(F)$ denotes the set of resource classes of the set of resources that appear in $F$. \textit{SLO types} are a well researched area \cite{beyer2018site,adoptslogoogle}. Common examples which seem to appear across different works in this area are latency, throughput, error/success rate, etc. These capture which kind of measurement one may want to do over a resource. SLO types are denoted by $S$. 

Typically $C$ is associated with the resource, and $S$ with the metric. So given a tuple $t = (r, m, A)$, $C(t) = C(r)$ and $S(t) = S(m)$. Therefore, we have the natural class association between the tuples $(r, m, A)$ and $(C(r), S(m))$. Using this, we introduce the \textit{extended} monitor tuple as $\{r_{i}, m_{i}, A_{i}, C(r_{i}), S(m_{i})\}_{1\leq i \leq k}$. We now define the class recommendation problem.

\noindent \textbf{Monitor Class Recommendation Problem} Given a service $S$, it's underlying resources $r(S)$ and dependencies $d(S)$, generate the sets $C(M(S)), S(M(S))$.

In the next section we discuss an empirical method to augment the monitor data with class information, thereby generating extended tuple data.
\section{Monitor Classes: An Empirical Study}
\label{sec:empiricalstudy}
This empirical study serves as a preliminary but essential investigation leading up to the development of a monitor recommendation framework. This study will address the following research questions:
\begin{itemize}
    \item \textbf{Coverage}: Which resource classes (and SLO types) cover most of the monitors in our dataset?
    \item \textbf{Co-occurrence}: Are there pairs of resource classes (or SLO types) that tend to always or never coexist together?
    \item \textbf{Structural Correlation}: Is there some correlation between resource classes and SLO types, which underscores structural interrelationships?
    \item \textbf{Functionality driven monitoring}: In what ways do service properties, such as functionality and dependencies, correlate with the monitor classes? 
\end{itemize}
 In this section we will first describe the methodology for generating resource class and SLO type sets with high enough coverage. We will then discuss the exact classes and what they mean in the cloud service ecosystem, and their distribution in our dataset. The last few subsections will then focus on various statistical correlation properties of these class/type labels. In particular, in the last subsection, we describe a simple recommendation model which predicts monitor classes based on just the structural data of a service viz. the dependencies and components.
\subsection{Methodology for mining the resource classes and SLO types}


We started with a simple pooling of all the triplet data (comprising of the $\{F,m,A\}_{M}$ triplet along with the associated monitor name) and tried clustering them using the tokens appearing in their names. After trying few variations of this approach, we concluded that the name strings themselves are not good for primary clustering since a bulk of tokens were corresponding to the environment and service specific signals which lead to cluster formation around those same principles. While the clustering we observed was still meaningful, it had no utility in helping engineers create/update monitors. We therefore needed an NLP approach which would discard all the standard non-functional tokens and which will utilize the complete context. We decided to harness the power of large language models (LLMs) and in-context learning. Ref.~\cite{hasanbeig2023allure} discusses in-context learning and systematic evaluation framework.


Our goal here is to filter out \textit{functionality} related signals in the monitor data, which does seem to be present in most of the monitors. To this end, we used ChatGPT (based on GPT3.5) to analyze the triplet data, $\{F,m,A\}_{M}$, and generate the descriptions of what the monitor was actually doing. This created a new augmented data set of quadruples, $\{F,m,A,D\}_{M}$ where $D$ is the text description of the specific triplet $\{F,m,A\}_{M}$'s functionality from a resource and metric perspective. These descriptions still contained a lot of environment and service signals, and therefore the when we tried clustering them, we still encountered heavy environmental influence, even though resource groups were starting to form.

The third and final step was to use ChatGPT itself to start labeling the data points, instead of relying on clustering and hoping functionality distills through. To this end, we needed a comprehensive list of resource and SLO classes. This set of class labels was created incrementally. We, along with expert advisors, manually went through a randomized sample of $\{F,m,A,D\}_{M}$ data and created a small set of SLO and resource classes which covered them. The manually validated data consists of 20 unique monitors per each resource and SLO class, which are randomly sampled with replacement. In total, we validated 260 monitors for resource classes and 180 monitors for SLO classes.


We computed precision, recall and F1 score over the manually evaluated dataset to measure the quality of responses generated. These metrics are computed for resource classes and SLO classes individually at a class level and an overall score is generated for all the classes combined. For all the 13 resource classes combined we observed an average precision of 0.9423, recall  0.9530 and F1 score 0.9476. Ram-memory, container and certificate are the 3 classes that achieved highest precision, recall and F1 score of 1. Other classes like service level and CPU have a high precision of 1 but a recall of 0.7142 and 0.9523 respectively, with an F1 score of 0.8333 and 0.9756. The highest recall of 1 is also observed for API, dependency, compute cluster, storage, IO and paging memory. With high recall, the F1 score of the above classes are 0.97, 0.97, 0.94, 0.97, 0.91 and 0.91. The precision, recall and F1 score are all 0.95 for cache memory. None of the above class obtained the lowest F1 score of 0.8095 with 0.85 precision and 0.7727 recall. 

The average F1 score for 9 SLO classes is 0.7967, with a precision of 0.7944 and recall 0.8311. Capacity, freshness and interruption rate have the highest recall 1, with an F1 score of 0.85, 0.82 and 0.85 respectively. Only one class, latency achieved the highest precision of 1, with a recall of 0.9 and F1 score 0.95, which is the best among all the SLO classes. Success rate and success rate QoS both achieve a 0.95 precision and an F1 score of 0.76 and 0.74 as their recall values are 0.6333 and 0.6129. The availability and throughput classes obtained an F1 score of 0.8717 and 0.7778. The lowest F1 score of 0.52 is obtained for others class, with a precision and recall of 0.5 and 0.5556. 

In conclusion, using in-context learning based approach coupled with open-coding and incremental label generation, we were able to generate resource class and SLO type labels with a very high level of overall accuracy. We will discuss these classes in detail in the next subsection.

\subsection{Resource classes and Service Level Objectives monitored by cloud services}
In this section, we first start by defining the Resource and SLO classes obtained from the data mining process described in previous section and then analyze their distribution at a both monitor and service level. We identified 13 major resource classes corresponding to majority of monitors in our dataset:
\begin{itemize}
    \item \textbf{Service-level} class include resources that are specific to an individual service. Examples include feature components offered by a service that other services can use, internal components such as functions or pipelines, etc.
    \item \textbf{API} is another prominent resource class for a cloud service. API is used to access data, functionality, or services provided by another software application or service. 
    \item \textbf{Dependency} class represents the external services and components that a service relies on for its proper functioning. Dependencies can include other services, databases, messaging systems, and other infrastructure components. 
    \item \textbf{CPU} resource class refers to the compute bandwidth available on a cluster node. It is typically monitored in real-time, and alerts are generated if the CPU usage exceeds certain thresholds. 
    \item \textbf{Compute cluster} class provides the computing power needed to run applications and services. It is typically composed of multiple virtual machines or instances that can be provisioned and scaled on demand. Monitoring a compute cluster involves tracking its performance, availability, and utilization.
    \item \textbf{Storage} resource class represents the physical and virtual storage devices used by the service to store data. Monitoring the storage resource class can help identify any issues with the storage devices or the storage infrastructure, such as disk failures, data corruption, or insufficient storage capacity.
    \item \textbf{Ram-memory} resource class represents the available RAM (volatile memory). Apart from the simple fact that insufficient memory can lead to slow response times or even service crashes, active monitoring of RAM usage can help identify any memory leaks or inefficient memory usage patterns.
    \item \textbf{Cache-memory} resource class represents a type of high-speed memory used by a cloud services to store frequently accessed data or instructions. Monitoring the cache memory usage can help identify any cache misses, freshness issues or inefficient cache usage patterns. 
    \item \textbf{Container} are a very popular mechanism used for deploying and running applications in a cloud environment, as they can be easily moved across different computing environments without requiring any changes to the underlying code. Containers act as encapsulations of resources and functionalities and can be thought as the most granular independently hostable unit. They are typically monitored for availability.
    \item \textbf{Certificates} are often used to secure communication channels and authenticate users, and guarantee compliance. Monitoring certificates includes tracking expiration dates, certificate revocation lists, and usage patterns.
    \item \textbf{IO} is also a machine level resource (like CPU and RAM) and covers machine level data bandwidth. Its monitoring is important to ensure that data is being transferred efficiently and without errors, as well as to detect and diagnose any performance issues that may be impacting the overall performance of a cloud service.
    \item \textbf{Paging memory} monitoring is important to ensure that paging operations are occurring efficiently and without errors. This includes monitoring paging rates, page faults, and other metrics related to memory usage and performance.
\end{itemize}
Any resources that are not covered in the above classes are captured in the catch-all class ``None-of-the-above''. Examples from this class include job execution, data sovereignty, network, etc. 
%

\begin{comment}
    \begin{figure}[htp]
    \centering
    \includegraphics[width=5cm]{plots/resource_frequency_sorted.jpg}
    \caption{Breakdown of Resource Classes}
\end{figure}
\end{comment}

\begin{figure}[h]
\centering
\vspace{-2mm}
\scalebox{.28}{\input{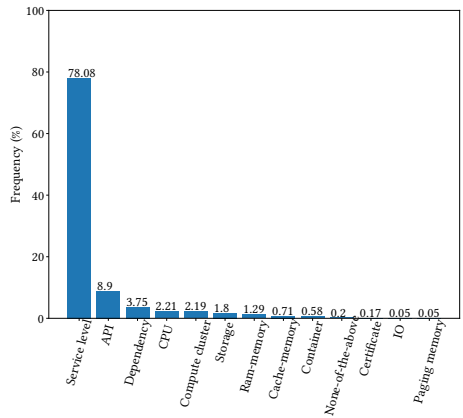}}
\vspace{-3mm}
    \caption{Breakdown of Resource Classes at Monitor Level. 
    }
\vspace{-2mm}
\label{fig:resource_classes}
\end{figure}

\begin{figure}[h]
\centering
\vspace{-2mm}
\scalebox{.22}{\input{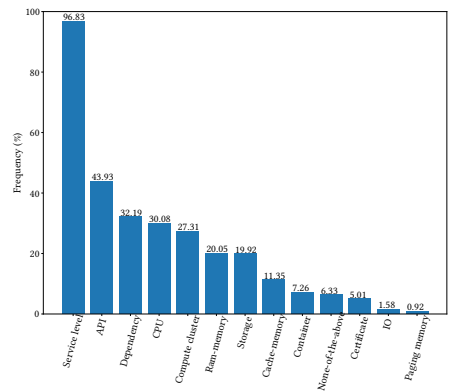}}
\vspace{-3mm}
    \caption{Breakdown of Resource Classes at Service Level. 
    }
\vspace{-2mm}
\label{fig:resource_classes_services}
\end{figure}

\autoref{fig:resource_classes} and \autoref{fig:resource_classes_services} summarize the distribution of major resource classes across monitors and services, respectively. In both the cases, the most popular resource classes being monitored ``Service-level'', ``API'', and ``Dependency", in that order. The popularity of ``Service level'' monitoring is intuitive as ensuring the service is meeting its various performance guarantees is the top priority of service owners. ``API'' is the second most popular resource type which is also easy to understand, since APIs are perhaps the most common design element in service oriented architecture. One final comment here is that it is evident that the number of ``Service level'' monitors within each service is relatively high compared to those for other resources. Since the ``Service level'' class includes features within a service and its internal components, we believe a finer granular classification of ``Service level'' monitors help with much better quality classification.

\begin{figure}[ht]
\centering
\vspace{-2mm}
\scalebox{.32}{\input{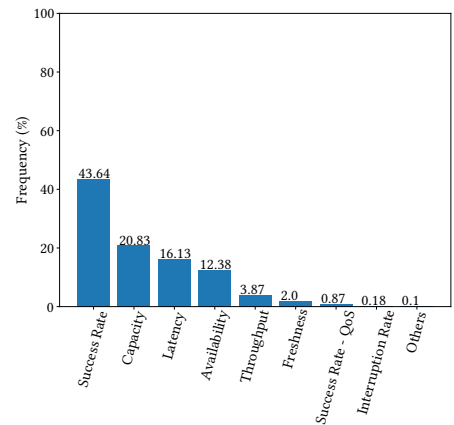}}
\vspace{-3mm}
    \caption{Breakdown of SLO Classes at Monitor Level
    }
\vspace{-2mm}
\label{fig:slo_classes}
\end{figure}
\begin{figure}[ht]
\centering
\vspace{-2mm}
\scalebox{.32}{\input{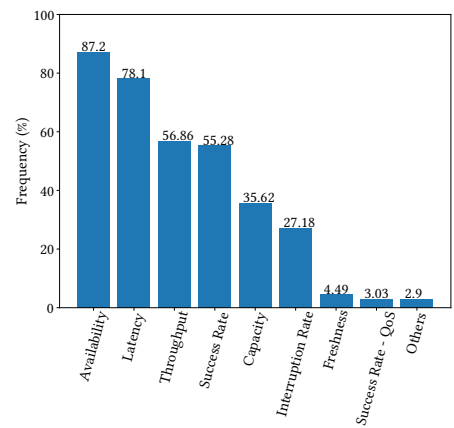}}
\vspace{-3mm}
    \caption{Breakdown of SLO Classes at Service Level}
\vspace{-2mm}
\label{fig:slo_classes_services}
\end{figure}

Next, we discuss the 9 major SLO classes obtained from the empirical data mining experiment. 
\begin{itemize}
    \item \textbf{Success rate} measures the number of successful events divided by the number of total events. For instance, ``99.99\% of requests in the last 5 minutes were successful". It is the most commonly used measurement type across monitors.
    \item \textbf{Capacity} SLOs have two types depending on whether capacity was actually available or not. In both cases the measurement is that of the number of throttling-based responses, in the first case it is counting the erroneous responses (since capacity is indeed available) and in second a genuine failure of meeting customer expectations.
    \item \textbf{Latency} is the amount of time elapsed between when a request for an operation is made and when the invoker can make use of the returned result.
    \item \textbf{Availability} measures the service's uptime, measured from the perspective of a customer trying to make use of the service, and is typically measured in 9's. Availability SLOs are similar to Success Rate SLOs but do not verify that the return results match what is expected of the requests, merely that return results are flowing to the user.
    \item \textbf{Throughput} SLOs measure a minimum data transfer rate over a time window, expressed in kilobytes per second. Throughput SLOs have highly variable time windows, dependent on the user expectation for when data transfer should finalize.
    \item \textbf{Success rate - QoS} focuses on the \textit{quality} of successful event.  It measures whether the service is performing as expected. E.g. reliability of a page rendered by the front end of a web app.
    \item \textbf{Interruption Rate} measures the rate of specific type of interruptive events on specific resource (e.g. the number of VM reboots.
\end{itemize}
Any metrics that are not covered in the above major classes are denoted by the class ``Others''. Examples from this class include compliance failures, privacy issues, etc.
\noindent \autoref{fig:slo_classes} summarizes the major SLO classes monitored by cloud services. The most popular SLO classes of monitors include ``Success Rate'', ``Capacity'' and ``Latency''. We then analyze the distribution of SLO classes across cloud services in \autoref{fig:slo_classes_services}. More than 50\% of cloud services monitor the ``Success Rate'', ``Capacity'', ``Availability'', and ``Latency'' of the  resources. ``Interruption Rate'' and ``Success Rate - QoS'' appear to be monitored by less than 5\% of services. This is intuitive since measuring interruptions is only applicable to certain resources, unlike classes such as ``Capacity''.

\begin{comment}
 \begin{figure}[htp]
    \centering
    \includegraphics[width=5cm]{plots/slo_frequency_sorted.jpg}
    \caption{Breakdown of SLO Classes}
\end{figure}   

\begin{figure}[ht]
\centering
\vspace{-2mm}
\scalebox{.23}{\input{plots/MonitorSlo.pgf}}%
\scalebox{.23}{\input{plots/Serviceslo_new.pgf}}
\vspace{-3mm}
    \caption{Breakdown of SLO Classes at Service Level}
\vspace{-2mm}
\label{fig:slo_classes_services}
\end{figure}
\end{comment}
Any metric that are not covered in the above major classes are denoted by the class ``Others''. Examples from this class include Compliance failures, privacy issues, etc.

\begin{takeaway}
Over 50\% of cloud services use monitors to measure success, capacity, availability, and latency. The widespread use of ``Service level'' monitors indicates that monitoring service levels is essential for ensuring that cloud services meet the needs of their users and applications. 
\end{takeaway}

\subsection{Major classes of Service Level Objectives associated with each resource class}
In this section, we analyze the association between SLO and resource classes.
\begin{figure}[ht]
\centering
\vspace{-2mm}
\scalebox{.32}{\input{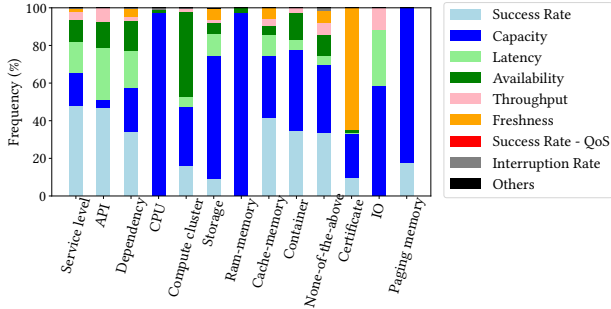}}
\vspace{-3mm}
    \caption{Distribution of SLO classes within each resource class}
\vspace{-3mm}
\label{fig:slo_resource_association}
\end{figure}
\autoref{fig:slo_resource_association}  illustrates the distribution of SLO classes within each resource class. Although ``Success Rate'' is overall the most commonly observed SLO class in the monitor data, the figure shows that it is not a leading member inside many individual resource classes. Evidently, the distribution of SLO classes varies across different resource classes. 

\myparagraph{Do resource classes define the critical SLO classes that apply to them?}
 To analyze  the distribution of  SLO classes across different resource classes with respect to the overall SLO distribution, we use Chi-squared test of independence \cite{pearson1900x} at a 5\% significance level (95\% confidence interval):

\noindent \textit{Null hypothesis} ($H_0$): The distribution of SLO classes inside resource classes is independent of the resource class itself.

\noindent \textit{Alternative hypothesis} ($H_A$): The distribution of SLO classes inside resource classes is a function of the resource class. 

We summarize the following insights for the cases where the overall distribution differs from those at the resource class level, i.e. the null hypothesis is rejected at 5\% significance level:
\begin{enumerate}
\item \textbf{API}, $H_0$ rejected with p-value = 0.0007. The most critical SLOs for resources in this class are \textbf{Success Rate} and \textbf{Latency}.
\item \textbf{Compute Cluster} class, $H_0$ rejected with p-value = 0. The key SLOs being monitored in this class are \textbf{Availability} and \textbf{Capacity}.
\item $H_0$ rejected for classes with corresponding p-values, \textbf{CPU} (p-value = 0), \textbf{Storage} (p-value = 0), \textbf{RAM memory} (p-value =0), \textbf{IO} (p-value = 0), \textbf{Container} (p-value = 0), and \textbf{Paging Memory} (p-value = 0). The predominant SLO to be monitored under all these classes is \textbf{Capacity}.
\item \textbf{Cache Memory}, $H_0$ rejected with p-value = 0.003. The key SLOs monitored for this class are \textbf{Success Rate} and \textbf{Capacity}.
\item \textbf{Certificate}, $H_0$ rejected with p-value = 0. The major SLO class for this class is \textbf{Freshness}.
\end{enumerate}

\begin{takeaway}
SLO class distribution varies across resource classes, suggesting a restricted set of metric classes to be considered for each class. Predicting a service's SLO classes can therefore be done by analyzing its associated resource classes.
\end{takeaway}

\subsection{Coexistence tendency among resource classes}
\begin{figure}[ht]
\centering
\vspace{-3mm}
\scalebox{.32}{\input{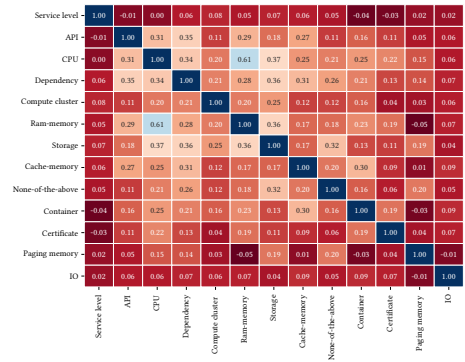}}
\vspace{-3mm}
    \caption{Coexistence of Resource Classes across Services}
\vspace{-2mm}
\label{fig:resource_correlation}
\end{figure}

Here, we examine whether certain resource class pairs have a higher than random, tendency to coexist and if the presence of one class implies the presence or absence of others.  We calculate the phi coefficient (mean square contingency coefficient, $\varphi$) to measure association between two binary variables \cite{hirschfeld1935connection}. The coefficient ranges from -1 to 1, with positive values indicating positive correlation and negative values indicating negative correlation. We only analyze coexistence among resource classes, as SLO class distribution depends heavily on underlying resource classes.
\autoref{fig:resource_correlation} shows the $\varphi$ coefficient between different resources. Correlation between \textbf{Service level} and \textbf{IO} monitors with other resource classes seems negligible. This is intuitive as \textbf{Service level} monitors depend on the functionality or internal components within the service. There is a strong correlation between \textbf{CPU} and \textbf{Ram-memory}. There is also a slight positive correlation among class pairs such as \textbf{API} and \textbf{Dependency}, \textbf{CPU} and \textbf{Storage}, \textbf{Dependency} and \textbf{Storage}, etc.
\begin{takeaway}
If the presence of a resource class within a service probabilistically implies the existence of another resource class, this signal can be utilized to suggest missing monitors for an existing service.
\end{takeaway}

\subsection{Service property similarities and resource class overlaps}
In this section, we analyze the predictability of ontology classes in relation to service properties. Since the distribution of SLO classes is influenced by the underlying resource classes, our focus is on the predictability of resource classes using service properties. Specifically, we consider two properties of a cloud service: 1) Dependencies associated with a service and 2) Physical and logical components within a service.

 \myparagraph{Service dependencies determining recourse classes}
In this section, we look at service dependencies, both upstream and downstream and examine the impact of service similarity in terms of their upstream and downstream dependencies when recommending monitors.
\begin{table}[h!]
\caption{Using upstream dependencies to recommend resource classes at service level (collaborative filtering)}
\centering
\resizebox{0.5\textwidth}{!}{%
\begin{tabular}{|c|p{2cm}|p{2cm}|p{2cm}|p{2cm}|p{2cm}|} 
 \hline
 Resource Type & Top 1 service & Top 2 services & Top 3 services & Top 4 services & Top 5 services \\
 \hline\hline
 service level & \begin{tabular}{@{}c@{}} AUC: 0.41\\Precision: 1.00\\Recall: 0.01 \end{tabular} & \begin{tabular}{@{}c@{}} AUC: 0.47\\Precision: 0.98\\Recall: 0.44 \end{tabular} & \begin{tabular}{@{}c@{}}AUC: 0.43\\Precision: 0.99\\Recall: 0.09 \end{tabular} & \begin{tabular}{@{}c@{}}AUC: 0.44\\Precision: 1.00\\Recall: 0.04 \end{tabular}& \begin{tabular}{@{}c@{}}AUC: 0.47\\Precision: 0.98\\Recall: 0.49 \end{tabular}\\
 \hline
 api & \begin{tabular}{@{}c@{}} AUC: 0.51\\Precision: 0.46\\Recall: 0.35 \end{tabular} & \begin{tabular}{@{}c@{}}AUC: 0.52\\Precision: 0.49\\Recall: 0.29 \end{tabular}& \begin{tabular}{@{}c@{}}AUC: 0.53\\Precision: 0.51\\Recall: 0.33 \end{tabular}& \begin{tabular}{@{}c@{}}AUC: 0.51\\Precision: 0.50\\Recall: 0.20 \end{tabular}& \begin{tabular}{@{}c@{}}AUC: 0.50\\Precision: 0.46\\Recall: 0.36 \end{tabular}\\
 \hline
 cpu & \begin{tabular}{@{}c@{}} AUC: 0.55\\Precision: 0.47\\Recall: 0.23 \end{tabular} &\begin{tabular}{@{}c@{}}AUC: 0.55\\Precision: 0.44\\Recall: 0.26 \end{tabular} & \begin{tabular}{@{}c@{}} AUC: 0.58\\Precision: 0.42\\Recall: 0.36\end{tabular}&\begin{tabular}{@{}c@{}}AUC: 0.56\\Precision: 0.40\\Recall: 0.33 \end{tabular} & \begin{tabular}{@{}c@{}}AUC: 0.55\\Precision: 0.40\\Recall: 0.27 \end{tabular}\\
 \hline
 container & \begin{tabular}{@{}c@{}} AUC: 0.57\\Precision: 0.36\\Recall: 0.18 \end{tabular} &\begin{tabular}{@{}c@{}}AUC: 0.56\\Precision: 0.24\\Recall: 0.16 \end{tabular} & \begin{tabular}{@{}c@{}} AUC: 0.55\\Precision: 0.27\\Recall: 0.15\end{tabular}&\begin{tabular}{@{}c@{}}AUC: 0.55\\Precision: 0.25\\Recall: 0.15 \end{tabular} &\begin{tabular}{@{}c@{}}AUC: 0.56\\Precision: 0.38\\Recall: 0.16 \end{tabular} \\
 \hline
 dependency & \begin{tabular}{@{}c@{}} AUC: 0.50\\Precision: 0.35\\Recall: 0.24 \end{tabular} &\begin{tabular}{@{}c@{}}AUC: 0.50\\Precision: 0.34\\Recall: 0.25 \end{tabular} &\begin{tabular}{@{}c@{}}AUC: 0.53\\Precision: 0.39\\Recall: 0.29 \end{tabular} & \begin{tabular}{@{}c@{}}AUC: 0.55\\Precision: 0.45\\Recall: 0.27 \end{tabular}&\begin{tabular}{@{}c@{}}AUC: 0.53\\Precision: 0.41\\Recall: 0.24 \end{tabular} \\
 \hline
 compute cluster & \begin{tabular}{@{}c@{}} AUC: 0.59\\Precision: 0.46\\Recall: 0.31 \end{tabular} &\begin{tabular}{@{}c@{}}AUC: 0.56\\Precision: 0.39\\Recall: 0.32 \end{tabular} &\begin{tabular}{@{}c@{}}AUC: 0.56\\Precision: 0.42\\Recall: 0.26 \end{tabular} &\begin{tabular}{@{}c@{}}AUC: 0.57\\Precision: 0.38\\Recall: 0.35 \end{tabular} &\begin{tabular}{@{}c@{}}AUC: 0.56\\Precision: 0.39\\Recall: 0.29 \end{tabular} \\
 \hline
 storage & \begin{tabular}{@{}c@{}} AUC: 0.56\\Precision: 0.35\\Recall: 0.23 \end{tabular} &\begin{tabular}{@{}c@{}}AUC: 0.55\\Precision: 0.32\\Recall: 0.22 \end{tabular} &\begin{tabular}{@{}c@{}}AUC: 0.53\\Precision: 0.34\\Recall: 0.12 \end{tabular} &\begin{tabular}{@{}c@{}}AUC: 0.53\\Precision: 0.45\\Recall: 0.10 \end{tabular} &\begin{tabular}{@{}c@{}}AUC: 0.50\\Precision: 0.35\\Recall: 0.06 \end{tabular} \\
 \hline
 ram-memory & \begin{tabular}{@{}c@{}} AUC: 0.56\\Precision: 0.35\\Recall: 0.23 \end{tabular} &\begin{tabular}{@{}c@{}}AUC: 0.54\\Precision: 0.38\\Recall: 0.18 \end{tabular} &\begin{tabular}{@{}c@{}} AUC: 0.55\\Precision: 0.31\\Recall: 0.24\end{tabular} &\begin{tabular}{@{}c@{}}AUC: 0.53\\Precision: 0.34\\Recall: 0.14 \end{tabular} &\begin{tabular}{@{}c@{}}AUC: 0.53\\Precision: 0.30\\Recall: 0.20 \end{tabular} \\
 \hline
 certificate & \begin{tabular}{@{}c@{}} AUC: 0.55\\Precision: 0.38\\Recall: 0.13 \end{tabular} &\begin{tabular}{@{}c@{}}AUC: 0.54\\Precision: 0.18\\Recall: 0.11 \end{tabular} &\begin{tabular}{@{}c@{}}AUC: 0.52\\Precision: 0.14\\Recall: 0.05 \end{tabular} &\begin{tabular}{@{}c@{}}AUC: 0.52\\Precision: 0.15\\Recall: 0.05 \end{tabular} &\begin{tabular}{@{}c@{}}AUC: 0.51\\Precision: 0.11\\Recall: 0.05 \end{tabular} \\
 \hline
\end{tabular}
}
\label{Tab:upstreamsimilarity}
\end{table}
\noindent To analyze the role of upstream services in monitor recommendation, we calculate the ``Top-n'' similar services for each service based on the cosine similarity among their upstream services one-hot encoded vector. Using the weighted sum of similarity among services and the normalized occurrence of each resource within the service as a score, we generate top recommendations: \\
Score$(S_i,C) = \sum_{S_j \in S(N)} d(S_i,S_j) N (C,{S_j})$ 
\\
Here, Score$(S_i,C)$  represents recommended score of resource class, $C$ for  service, $S_i$. Scalar, $d(S_i,S_j)$ represents cosine distance between $S_i$ and $S_j$ in their feature space, $N_{C\ \times {S_j}}$ is a matrix representing the resource classes within ${S_j}$'s, and $S(N)$  is a set of top $n$ similar services to $S_i$, which appear in the matrix $N$. We follow the same approach for generating score for other features.
\autoref{Tab:upstreamsimilarity} presents the precision, recall, and AUC of the predictions based on Top 1-5 services. We use Youden's Index to determine the optimal threshold for the predictions \cite{hilden1996regret}.

\noindent Similarly, we generate recommendations using downstream services in the service graph for a given service. The precision, recall, and AUC of the recommendations based on Top 1-5 services are summarized in \autoref{Tab:downstreamsimilarity}.
\begin{table}[h!]
\caption{Using downstream dependencies to recommend resource classes at service level (collaborative filtering)}
\centering
\resizebox{0.5\textwidth}{!}{%
\begin{tabular}{|c|p{2cm}|p{2cm}|p{2cm}|p{2cm}|p{2cm}|} 
 \hline
 Resource Type & Top 1 service & Top 2 services & Top 3 services & Top 4 services & Top 5 services \\
 \hline\hline
 service level & \begin{tabular}{@{}c@{}}AUC: 0.46\\Precision: 1.00\\Recall: 0.07  \end{tabular} & \begin{tabular}{@{}c@{}} AUC: 0.40\\Precision: 1.00\\Recall: 0.01 \end{tabular} & \begin{tabular}{@{}c@{}}AUC: 0.45\\Precision: 1.00\\Recall: 0.10 \end{tabular} & \begin{tabular}{@{}c@{}}AUC: 0.45\\Precision: 1.00\\Recall: 0.09 \end{tabular}& \begin{tabular}{@{}c@{}}AUC: 0.47\\Precision: 1.00\\Recall: 0.07 \end{tabular}\\
 \hline
 api & \begin{tabular}{@{}c@{}}AUC: 0.50\\Precision: 0.61\\Recall: 0.06  \end{tabular} & \begin{tabular}{@{}c@{}}AUC: 0.51\\Precision: 0.53\\Recall: 0.20 \end{tabular}& \begin{tabular}{@{}c@{}}AUC: 0.48\\Precision: 1.00\\Recall: 0.01 \end{tabular}& \begin{tabular}{@{}c@{}}AUC: 0.50\\Precision: 0.56\\Recall: 0.18 \end{tabular}& \begin{tabular}{@{}c@{}}AUC: 0.51\\Precision: 0.56\\Recall: 0.20 \end{tabular}\\
 \hline
 cpu & \begin{tabular}{@{}c@{}}AUC: 0.50\\Precision: 0.48\\Recall: 0.08  \end{tabular} &\begin{tabular}{@{}c@{}}AUC: 0.52\\Precision: 0.43\\Recall: 0.14 \end{tabular} & \begin{tabular}{@{}c@{}} AUC: 0.51\\Precision: 0.38\\Recall: 0.21\end{tabular}&\begin{tabular}{@{}c@{}}AUC: 0.50\\Precision: 0.39\\Recall: 0.12 \end{tabular} & \begin{tabular}{@{}c@{}}AUC: 0.51\\Precision: 0.42\\Recall: 0.15 \end{tabular}\\
 \hline
 container & \begin{tabular}{@{}c@{}}AUC: 0.51\\Precision: 0.15\\Recall: 0.07  \end{tabular} &\begin{tabular}{@{}c@{}}AUC: 0.50\\Precision: 1.00\\Recall: 0.02 \end{tabular} & \begin{tabular}{@{}c@{}} AUC: 0.51\\Precision: 1.00\\Recall: 0.02\end{tabular}&\begin{tabular}{@{}c@{}}AUC: 0.50\\Precision: 1.00\\Recall: 0.02 \end{tabular} &\begin{tabular}{@{}c@{}}AUC: 0.51\\Precision: 1.00\\Recall: 0.02 \end{tabular} \\
 \hline
 dependency  & \begin{tabular}{@{}c@{}}AUC: 0.50\\Precision: 0.39\\Recall: 0.11  \end{tabular} &\begin{tabular}{@{}c@{}}AUC: 0.50\\Precision: 0.43\\Recall: 0.14 \end{tabular} & \begin{tabular}{@{}c@{}} AUC: 0.50\\Precision: 0.42\\Recall: 0.11\end{tabular}&\begin{tabular}{@{}c@{}} AUC: 0.50\\Precision: 0.38\\Recall: 0.18\end{tabular} &\begin{tabular}{@{}c@{}}AUC: 0.51\\Precision: 0.50\\Recall: 0.10 \end{tabular} \\
 \hline
 compute cluster  & \begin{tabular}{@{}c@{}}AUC: 0.55\\Precision: 0.46\\Recall: 0.19  \end{tabular} &\begin{tabular}{@{}c@{}}AUC: 0.52\\Precision: 0.35\\Recall: 0.12 \end{tabular} & \begin{tabular}{@{}c@{}}AUC: 0.52\\Precision: 0.56\\Recall: 0.07 \end{tabular}&\begin{tabular}{@{}c@{}}AUC: 0.53\\Precision: 0.54\\Recall: 0.09 \end{tabular} &\begin{tabular}{@{}c@{}}AUC: 0.50\\Precision: 0.67\\Recall: 0.01 \end{tabular} \\
 \hline
 storage  & \begin{tabular}{@{}c@{}}AUC: 0.50\\Precision: 0.43\\Recall: 0.05  \end{tabular} &\begin{tabular}{@{}c@{}}AUC: 0.51\\Precision: 0.44\\Recall: 0.07 \end{tabular} & \begin{tabular}{@{}c@{}} AUC: 0.47\\Precision: 0.38\\Recall: 0.03\end{tabular}&\begin{tabular}{@{}c@{}}AUC: 0.49\\Precision: 0.26\\Recall: 0.08 \end{tabular} &\begin{tabular}{@{}c@{}}AUC: 0.52\\Precision: 0.42\\Recall: 0.13 \end{tabular} \\
 \hline
 ram-memory  & \begin{tabular}{@{}c@{}}AUC: 0.51\\Precision: 0.31\\Recall: 0.08  \end{tabular} &\begin{tabular}{@{}c@{}}AUC: 0.51\\Precision: 0.30\\Recall: 0.09 \end{tabular} & \begin{tabular}{@{}c@{}} AUC: 0.49\\Precision: 0.36\\Recall: 0.04\end{tabular}&\begin{tabular}{@{}c@{}}AUC: 0.51\\Precision: 0.29\\Recall: 0.12 \end{tabular} &\begin{tabular}{@{}c@{}}AUC: 0.49\\Precision: 0.00\\Recall: 0.00 \end{tabular} \\
 \hline
 certificate  & \begin{tabular}{@{}c@{}}AUC: 0.53\\Precision: 0.67\\Recall: 0.07  \end{tabular} &\begin{tabular}{@{}c@{}}AUC: 0.51\\Precision: 0.50\\Recall: 0.04 \end{tabular} & \begin{tabular}{@{}c@{}}AUC: 0.51\\Precision: 0.50\\Recall: 0.04 \end{tabular}&\begin{tabular}{@{}c@{}}AUC: 0.51\\Precision: 0.50\\Recall: 0.04 \end{tabular} &\begin{tabular}{@{}c@{}}AUC: 0.51\\Precision: 0.50\\Recall: 0.04 \end{tabular} \\
 \hline
\end{tabular}
}
\label{Tab:downstreamsimilarity}
\end{table}
\begin{takeaway}
The resource classes to be monitored for a given service are influenced by the upstream and downstream services within the service graph, with the exception of the ``service-level" monitors. Upstream services in general influence the resource class to be monitored compared to downstream services. 
\end{takeaway}
\myparagraph{Do services with similar components employ similar monitors?}
In this section, we analyze the correlation between the components associated with a service and their implications on where do add monitors for those services. These components can be physical components such as Datacenters or logical components like a Role. Cloud provider have access to these components associated with the service and we look at the similarity between services based on these components, identify similar services, and  provide recommendations.  \autoref{tab:dim_recco} shows the precision, recall, and AUC of the predictions with n (number of similar services) varying from one to five. 

Based on the results from \autoref{Tab:upstreamsimilarity}, \ref{Tab:downstreamsimilarity}, and \ref{tab:dim_recco}, we can observe that each class's prediction is influenced by different features to varying degrees. Further, number of Top services for better predictions also varies with resource classes and service properties. The AUC scores and precision values reflect this. For example, components are the most useful feature for predicting the \textbf{Compute cluster}, \textbf{Certificate}, and \textbf{API} classes, while upstream dependency is the most useful feature for the \textbf{Dependency} class. Similarly, recommendations based on Top 5 similar services calculated based on upstream dependencies show better results for \textbf{Dependency} class.

\begin{table}[h!]
\caption{Recommendations using service-based collaborative filtering (similar services are identified based on the service architecture).}
\centering
\resizebox{0.5\textwidth}{!}{%
\begin{tabular}{|c|p{2cm}|p{2cm}|p{2cm}|p{2cm}|p{2cm}|} 
 \hline
 Resource Type & Top 1 service & Top 2 services & Top 3 services & Top 4 services & Top 5 services \\
 \hline\hline
 service level & \begin{tabular}{@{}c@{}}AUC: 0.47\\Precision: 0.98\\Recall: 0.27  \end{tabular} & \begin{tabular}{@{}c@{}}AUC: 0.41\\Precision: 0.99\\Recall: 0.1
 \end{tabular} & \begin{tabular}{@{}c@{}}AUC: 0.39\\Precision: 0.97\\Recall: 0.29 \end{tabular} & \begin{tabular}{@{}c@{}}AUC: 0.38\\Precision: 0.98\\Recall: 0.12 \end{tabular}& \begin{tabular}{@{}c@{}}AUC: 0.38\\Precision: 0.98\\Recall: 0.14 \end{tabular}\\
 \hline
 api & \begin{tabular}{@{}c@{}}AUC: 0.54\\Precision: 0.64\\Recall: 0.15  \end{tabular} & \begin{tabular}{@{}c@{}}AUC: 0.54\\Precision: 0.62\\Recall: 0.17  \end{tabular} & \begin{tabular}{@{}c@{}}AUC: 0.54\\Precision: 0.75\\Recall: 0.14 \end{tabular} & \begin{tabular}{@{}c@{}}AUC: 0.54\\Precision: 0.78\\Recall: 0.14 \end{tabular}& \begin{tabular}{@{}c@{}}AUC: 0.53\\Precision: 0.74\\Recall: 0.14 \end{tabular}\\
 \hline
 cpu & \begin{tabular}{@{}c@{}}AUC: 0.53\\Precision: 0.78\\Recall: 0.1  \end{tabular} & \begin{tabular}{@{}c@{}}AUC: 0.53\\Precision: 0.69\\Recall: 0.12 \end{tabular} & \begin{tabular}{@{}c@{}}
 AUC: 0.53\\Precision: 0.64\\Recall: 0.12
 \end{tabular} & \begin{tabular}{@{}c@{}}
 AUC: 0.52\\Precision: 0.65\\Recall: 0.13
 \end{tabular}& \begin{tabular}{@{}c@{}}
AUC: 0.51\\Precision: 0.64\\Recall: 0.13
\end{tabular}\\
 \hline
 container & \begin{tabular}{@{}c@{}}
 AUC: 0.55\\Precision: 0.50\\Recall: 0.1
 \end{tabular} & \begin{tabular}{@{}c@{}}
AUC: 0.55\\Precision: 0.43\\Recall: 0.1
 \end{tabular} & \begin{tabular}{@{}c@{}}
AUC: 0.55\\Precision: 0.43\\Recall: 0.1
\end{tabular} & \begin{tabular}{@{}c@{}}
AUC: 0.55\\Precision: 0.43\\Recall: 0.1
\end{tabular}& \begin{tabular}{@{}c@{}}
AUC: 0.55\\Precision: 0.43\\Recall: 0.1
\end{tabular}\\
 \hline
 dependency  & \begin{tabular}{@{}c@{}}
  AUC: 0.51\\Precision: 0.36\\Recall: 0.05
  \end{tabular} & \begin{tabular}{@{}c@{}}
   AUC: 0.52\\Precision: 0.32\\Recall: 0.06
   \end{tabular} & \begin{tabular}{@{}c@{}}
AUC: 0.52\\Precision: 0.27\\Recall: 0.06
   \end{tabular} & \begin{tabular}{@{}c@{}}
AUC: 0.52\\Precision: 0.3\\Recall: 0.07 
   \end{tabular}& \begin{tabular}{@{}c@{}}
AUC: 0.52\\Precision: 0.32\\Recall: 0.06
 \end{tabular}\\
 \hline
 compute cluster & \begin{tabular}{@{}c@{}}
 AUC: 0.59\\Precision: 0.75\\Recall: 0.21
 \end{tabular} & \begin{tabular}{@{}c@{}}
AUC: 0.6\\Precision: 0.73\\Recall: 0.23 \end{tabular} & \begin{tabular}{@{}c@{}}
AUC: 0.61\\Precision: 0.71\\Recall: 0.24
 \end{tabular} & \begin{tabular}{@{}c@{}}
AUC: 0.61\\Precision: 0.7\\Recall: 0.25
\end{tabular}& \begin{tabular}{@{}c@{}}
AUC: 0.61\\Precision: 0.7\\Recall: 0.25
 \end{tabular}\\
 \hline
 storage & \begin{tabular}{@{}c@{}}
 AUC: 0.53\\Precision: 0.64\\Recall: 0.08
 \end{tabular} & \begin{tabular}{@{}c@{}}
AUC: 0.53\\Precision: 0.85\\Recall: 0.06 \end{tabular} & \begin{tabular}{@{}c@{}}
AUC: 0.52\\Precision: 0.73\\Recall: 0.06
\end{tabular} & \begin{tabular}{@{}c@{}}
AUC: 0.52\\Precision: 0.63\\Recall: 0.07
 \end{tabular}& \begin{tabular}{@{}c@{}}
AUC: 0.52\\Precision: 0.6\\Recall: 0.07
 \end{tabular}\\
 \hline
 ram-memory & \begin{tabular}{@{}c@{}}AUC: 
 AUC: 0.55\\Precision: 0.53\\Recall: 0.12
 \end{tabular} & \begin{tabular}{@{}c@{}}
AUC: 0.56\\Precision: 0.43\\Recall: 0.16
\end{tabular} & \begin{tabular}{@{}c@{}}
AUC: 0.56\\Precision: 0.38\\Recall: 0.17
\end{tabular} & \begin{tabular}{@{}c@{}}
AUC: 0.55\\Precision: 0.46\\Recall: 0.14
 \end{tabular}& \begin{tabular}{@{}c@{}}
AUC: 0.54\\Precision: 0.44\\Recall: 0.14
\end{tabular}\\
 \hline
 certificate & \begin{tabular}{@{}c@{}}
AUC: 0.54\\Precision: 0.75\\Recall: 0.09
 \end{tabular} & \begin{tabular}{@{}c@{}}
AUC: 0.55\\Precision: 0.79\\Recall: 0.11 \end{tabular} & \begin{tabular}{@{}c@{}}
AUC: 0.56\\Precision: 0.80\\Recall: 0.12
\end{tabular} & \begin{tabular}{@{}c@{}}
AUC: 0.55\\Precision: 0.71\\Recall: 0.12
 \end{tabular}& \begin{tabular}{@{}c@{}}
AUC: 0.55\\Precision: 0.71\\Recall: 0.12
 \end{tabular}\\
 \hline
\end{tabular}
}
\label{tab:dim_recco}
\end{table}

\begin{takeaway}
The physical and logical components within a cloud service impact the resource classes that need to be monitored, with the exception of the ``Service level''. The predictability of the ``Service level'' monitor is low based on all the service properties, which indicates a need for subclasses. 
\end{takeaway}
\begin{comment}
\subsection{Summary}
\begin{itemize}
\item Distribution of SLO classes depends on the underlying resource class from the cloud service. Therefore, if we recommend resource classes, the SLO classes will follow
\item Certain resource classes tend to coexist together and the presence of existing resource classes is a good feature while recommending missing monitors for existing services.
\item ``Where to put monitors'' within a cloud service depends on the service properties.
Baseline  model for recommending  ``where to add'' a monitor can be constructed with the dependency graph of a service and the set of components within a service.
\item Predictability of service level monitor is low and indicates need for subclasses within. 
\end{itemize}  
\end{comment}
\section{Intelligent Monitor Recommendation Framework}
Section~\ref{sec:empiricalstudy} presents two key observations. Firstly, there exists a small set of SLO and resource classes which \textit{describe} almost all the monitors. Secondly, service properties have signals which are helpful in determining ``what'' to monitor on a cloud service. In this section, we build upon these insights to develop a more comprehensive intelligent monitoring framework 
for a new service. We begin by defining the overall framework, model design, and finally, present the experiments and evaluation.
\subsection{Framework}
The goal of this recommendation framework is to address the \textbf{Monitor Class Recommendation Problem} outlined in section \ref{sec:background}. Following the conclusion in the last section, we will focus only on inferring resource classes for monitors in new service scenarios i.e. recommending $C(M(S))$ for a new service $S$. 

For all service level data points, the following one-hot encoding based features are generated: 1) upstream dependency, 2) downstream dependency, and 3) components. Using these as base representations of services, we learn their associations with the resource classes. This would typically be posed as a multi-label learning problem, where in every service is eligible to receive multiple labels (resource classes) assigned to it, with varying levels of confidence. Since the label set size is small, we have split the multi-label classification problem into individual resource class prediction problems. These models are trained and evaluated separately and we discuss them in {\autoref{sec:evaluation}}.
\subsection{Model Design}
 We denote the training dataset by $D=\{(\bold{x}_i,y_i) \}_{i=1}^{n}$ with $\bold{x}_i \in \mathbb{R}^p$ and $y_i \in \{0,1\}$ for each $ i \in \{1, \ldots, n \}$. Based on \autoref{sec:empiricalstudy}, all the service properties and recommendations based on top one to five similar services are not relevant for each resource class. We need to select features that are relevant based on the class we are going to predict. To this end, we need a distance based classifier in a low-dimensional learned feature space. To learn patterns from the data, we use the Prototypical learning network proposed by Li et al.~\cite{li2018deep}.  The model
architecture consists of two components: an autoencoder (in
cluding an encoder, $f:\mathbb{R}^{p} \rightarrow \mathbb{R}^q$ and a decoder, $g: \mathbb{R}^q \rightarrow \mathbb{R}^p$ ) and a prototype classification network ($h: \mathbb{R}^q \rightarrow \mathbb{R}^K$). The network uses the autoencoder to reduce the dimensionality of the input and to learn useful features for prediction; then it uses the encoded input to produce a probability distribution over the $k$ classes through the prototype classification network $h$. The network $h$ is made up of three layers: a prototype layer, $p: \mathbb{R}^q \rightarrow \mathbb{R}^m$, a fully-connected layer $w : \mathbb{R}^m \rightarrow \mathbb{R}^k$, and a softmax layer, $s: \mathbb{R}^K \mathbb{R}^K$. The network learns $m$ prototype vectors $\mathbf{p}_1, \ldots, \mathbf{p}_m \in \mathbb{R}^q$ (each corresponds to a prototype unit) in the latent space. The prototype layer p
 computes the squared $\mathcal{L}^2$ distance between the encoded input $\mathbf{z} = f(\mathbf{x}_i)$ and each of the prototype vectors:  $ p(\mathbf{z}) = \left[ \|\mathbf{z} -p_1\|_2^2, ~ \|\mathbf{z} -p_2\|_2^2, ~\ldots,~ \|\mathbf{z} -p_m\|_2^2\right]$.
  \noindent The prototype unit corresponding to $p_j$ executes the computation $\|\mathbf{z} - p_j\|_2^2$. The fully-connected layer $w$
 computes weighted sums of these distances $Wp(\mathbf{z})$, where W is a $K \times m$ weight matrix. These weighted sums are then
 normalized by the softmax layer $s$ to output a probability
 distribution over the K classes. 
 During prediction, the model outputs the class that it thinks is the most probable. 
 The cost function consists of a standard cross-entropy term that penalizes misclassification, as well as terms that encourage learned prototypes to be similar to meaningful points from input vectors.
 
\noindent This approach provides two advantages:
1. It can automatically learn relevant features and prototypes that are representative of each class.
2. The prototype vectors live in the same space as the encoded inputs, which allows us to leverage these vectors, feed them into the decoder, and visualize the learned prototypes. This property can be used to interpret the results and display the key features behind the given recommendations, increasing explainability to the engineers.

\begin{figure}
    \centering
    \includegraphics[scale=0.25]{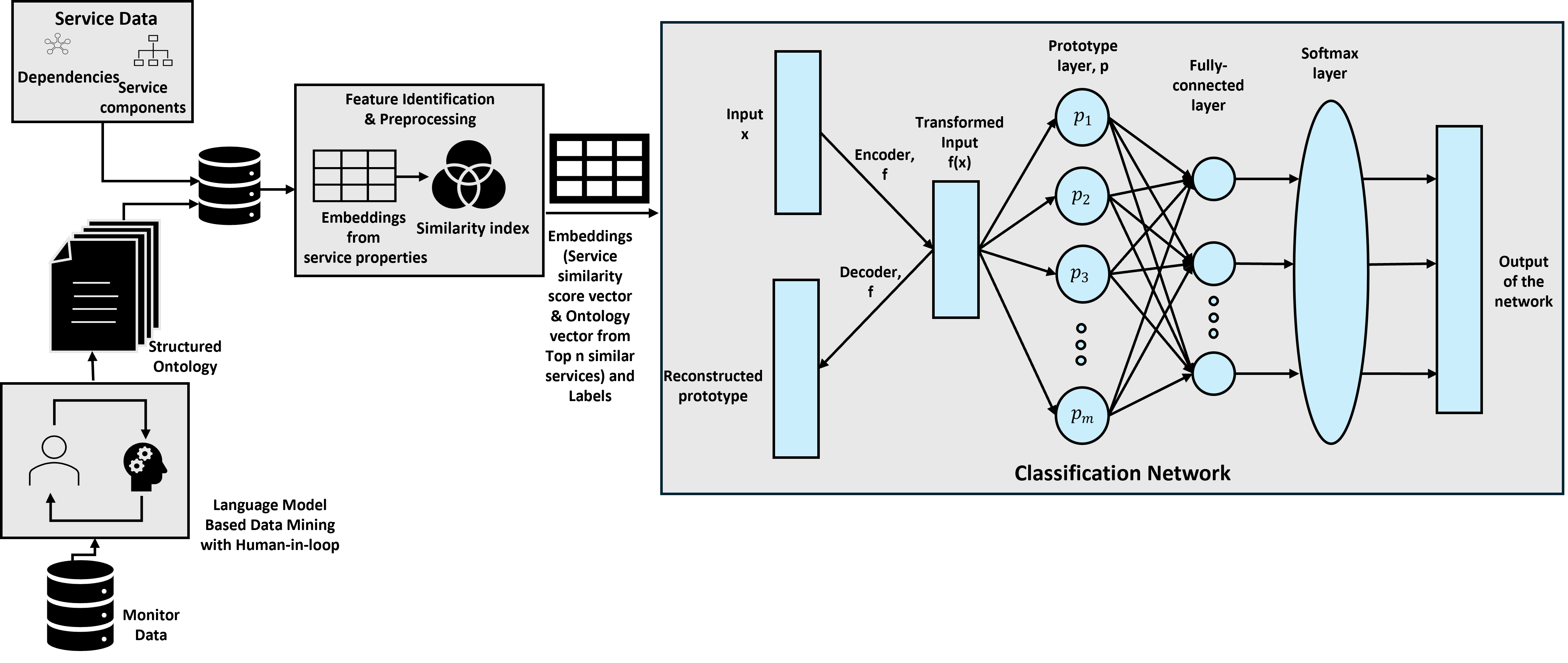}
     \vspace{-0.15in}
    \caption{Overview of the Recommendation Pipeline}
    \vspace{-0.15in}
    \label{fig:framework}
\end{figure}

\subsection{Experiments and Evaluation} 
\label{sec:evaluation}
The dataset we use for baseline recommendation comprises of normalised representation of resource classes for each service, along with the similarity matrices for upstream dependencies, downstream dependencies and components of the service. We evaluate predictions only for those resource classes where percentage of services that employ that class is greater than 5\%. Using the similarity score from Top 5 services and corresponding resource class matrix, the input to the model, $x$ is of dimension $\mathbb{R}^5$. The 791 services in our dataset are split in an 80:20 ratio to create a train dataset with 606 samples and a test dataset with 152 samples. To achieve equal representation of positive and negative samples for each resource, we perform upsampling on only the train dataset. The \textbf{Service level} class, consists of 587 positive samples, with the most skewed representation across classes. The train dataset for service level is upsampled to 1178 samples to account for the difference in representations. Other classes, such as \textbf{API}, \textbf{Dependency}, \textbf{CPU}, and \textbf{Compute cluster} on an average have 200-400 positive and negative samples in the train dataset. The original train dataset of \textbf{Ram-memory}, \textbf{Cache-memory} and \textbf{Storage} contains 100 positive samples on an average, and hence the train dataset is upsampled to approximately 1000 samples to generate a balanced dataset. \textbf{Container}, \textbf{Certificate} and \textbf{None-of-the-above} are classes with lowest representation of less than 50 positive samples originally and are then up-sampled to contain over 1100 samples in the train dataset.  

\begin{figure}
    \centering
    \includegraphics[scale=0.25]{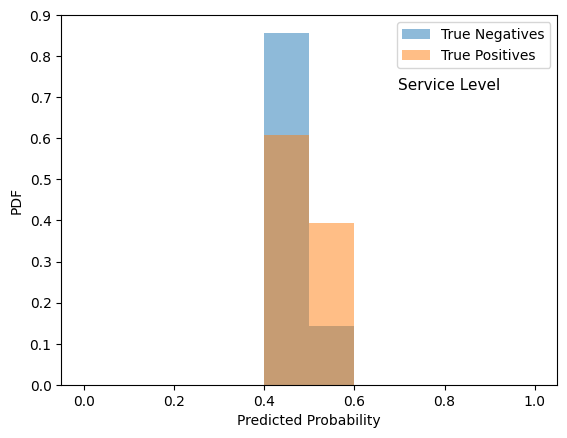}%
        \includegraphics[scale=0.25]{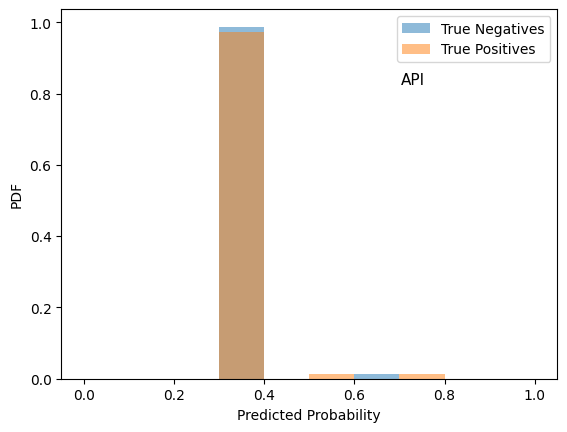}
  \includegraphics[scale=0.25]{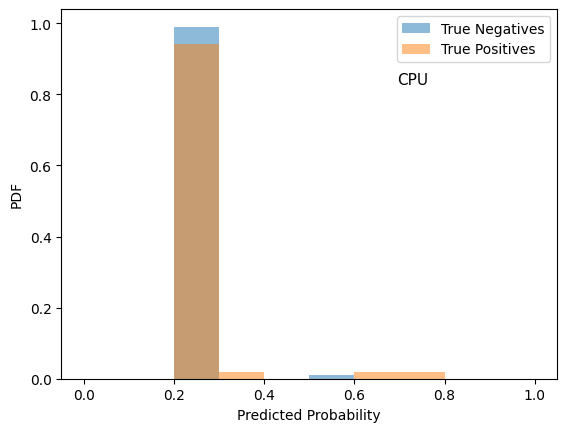}%
        \includegraphics[scale=0.25]{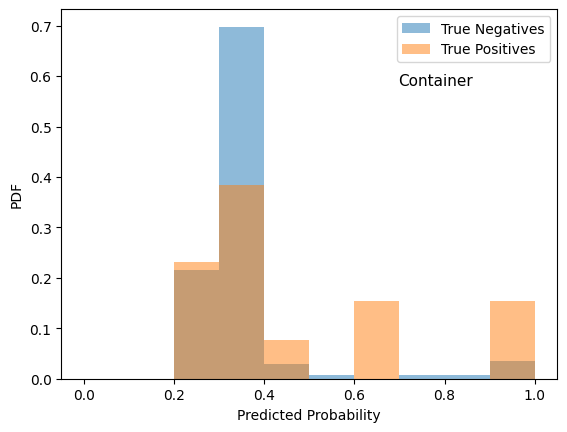}
 \includegraphics[scale=0.25]{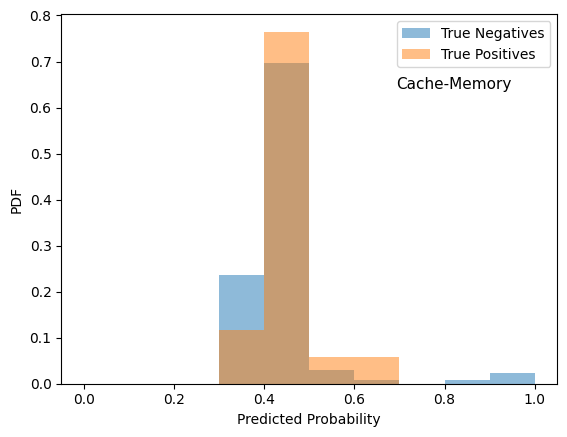}%
        \includegraphics[scale=0.25]{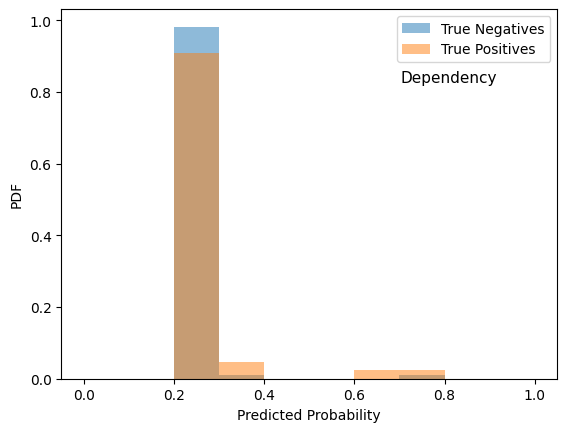}
         \includegraphics[scale=0.25]{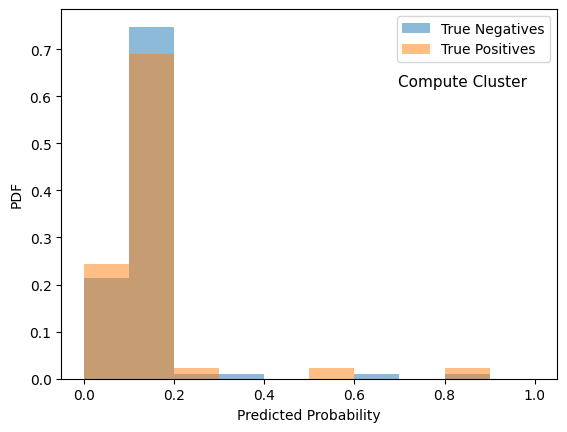}%
        \includegraphics[scale=0.25]{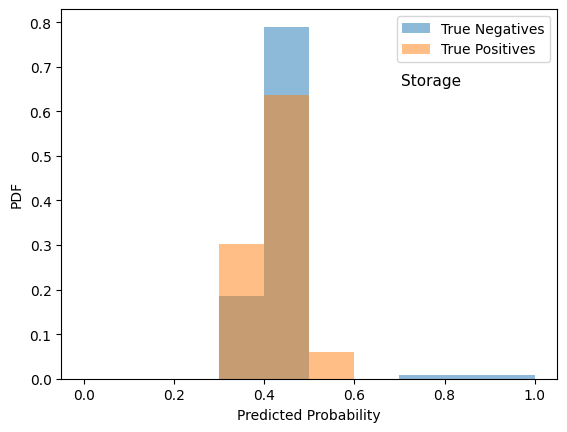}
                 \includegraphics[scale=0.25]{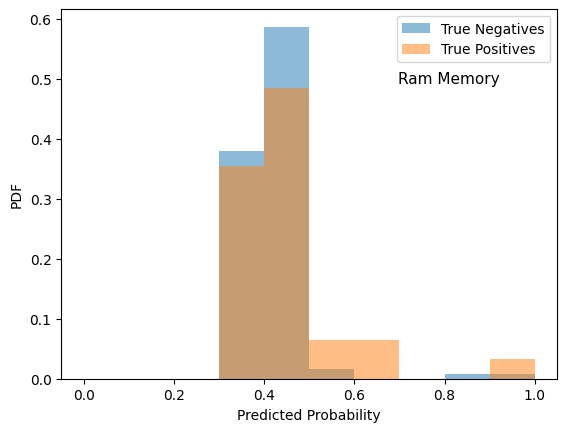}%
        \includegraphics[scale=0.25]{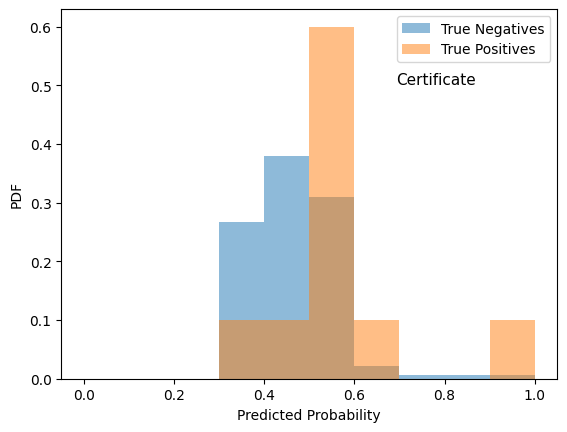}
        \includegraphics[scale=0.25]{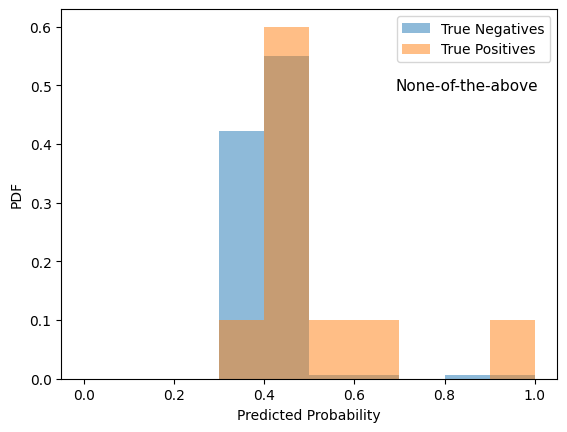}
     \vspace{-0.15in}
 \caption{Probability density of true positives and negatives for (a) Service Level,  (b) API, (c) CPU, (d) Container, (e) Cache-memory, (f) Dependency, (g) Compute Cluster, (h)  Storage,  (i) Ram-memory, (j) Certificate, (k) None-of-the-above classes. 
 } 
    \label{fig:PDF}
\end{figure}

\myparagraph{How effective is the recommendation framework in suggesting  monitors for new  services?}
 \autoref{fig:PDF} illustrates the density of positive and negative populations with respect to the probability predictions from the model. By setting custom thresholds, most of these classes generate production-quality recommendations, and Table \ref{tab:finalrecco} illustrates the quantitative metrics for these classes. In general, the results include three cases: 1) overall positive density is very high, making the prediction task nearly trivial; 2) positive density is low, but positive and negative density separation is clear. Examples of this are \textbf{Container} and \textbf{Certificate}; and 3) density is also low, and separation has not occurred well. This is the case where the ontology and the model have not created enough signals to make meaningful predictions.
\begin{table}[h!]
\caption{Quantitative metrics evaluated on recommendations from the proposed framework.}
\centering
\resizebox{0.5\textwidth}{!}{%
\begin{tabular}{|c|p{2cm}|p{2cm}|p{2cm}|} 

 \hline
 Resource Class & Threshold & Precision & Recall \\
 \hline\hline
 Service level & \begin{tabular}{@{}c@{}}0.45 \end{tabular} & \begin{tabular}{@{}c@{}}0.95 \end{tabular} & \begin{tabular}{@{}c@{}}1.00 \end{tabular}\\
 \hline
 API & \begin{tabular}{@{}c@{}} 0.30\\\end{tabular} & \begin{tabular}{@{}c@{}}0.48 \end{tabular} & \begin{tabular}{@{}c@{}} 1.00\\\end{tabular}\\
 \hline
 CPU & \begin{tabular}{@{}c@{}}0.20 \end{tabular} & \begin{tabular}{@{}c@{}}0.34 \end{tabular} & \begin{tabular}{@{}c@{}}1.00 \end{tabular}\\
 \hline
 Container & \begin{tabular}{@{}c@{}}0.40\end{tabular} & \begin{tabular}{@{}c@{}}0.30 \end{tabular} & \begin{tabular}{@{}c@{}}0.38 \end{tabular}\\
 \hline
 Dependency  & \begin{tabular}{@{}c@{}}0.20 \end{tabular} & \begin{tabular}{@{}c@{}}0.28 \end{tabular} & \begin{tabular}{@{}c@{}} 1.00 \end{tabular}\\
 \hline
 Compute cluster & \begin{tabular}{@{}c@{}}0.05 \end{tabular} & \begin{tabular}{@{}c@{}}0.30 \end{tabular} & \begin{tabular}{@{}c@{}}1.00 \end{tabular}\\
 \hline
 Storage & \begin{tabular}{@{}c@{}}0.35 \end{tabular} & \begin{tabular}{@{}c@{}}0.22 \end{tabular} & \begin{tabular}{@{}c@{}}1.00 \end{tabular}\\
 \hline
 Ram-memory & \begin{tabular}{@{}c@{}}0.30 \end{tabular} & \begin{tabular}{@{}c@{}}0.20 \end{tabular} & \begin{tabular}{@{}c@{}}1.00 \end{tabular}\\
 \hline
 Certificate & \begin{tabular}{@{}c@{}}0.50 \end{tabular} & \begin{tabular}{@{}c@{}}0.14 \end{tabular} & \begin{tabular}{@{}c@{}}0.80
 \end{tabular}\\
 \hline
 Cache-memory & \begin{tabular}{@{}c@{}}0.41 \end{tabular} & \begin{tabular}{@{}c@{}}0.13 \end{tabular} & \begin{tabular}{@{}c@{}}0.88 \end{tabular}\\
 \hline
 None-of-the-above & \begin{tabular}{@{}c@{}} 0.40\end{tabular} & \begin{tabular}{@{}c@{}}0.10	 \end{tabular} & \begin{tabular}{@{}c@{}}0.90 \end{tabular}\\
 \hline
\end{tabular}
}
\label{tab:finalrecco}
\end{table}


Next, we summarize the takeaways from the evaluations: 1) We need to define finer sub-classes for \textbf{Service-level} resources, for which the positive density is relatively high; and 2) We need to find better features for classes with low positive density so that they can perform high-fidelity predictions.
\normalsize

\section{User Study}
We demonstrated the efficiency of the recommendation framework using quantitative metrics in \autoref{sec:evaluation}. Next, we conduct a user study to understand the importance and usefulness of the monitor recommendation and how engineers perceive the recommendations.

\myparagraph{Setup} To select study participants, we generated a list of engineers at \CompanyX{} who modified monitors between January and June 2023. We filtered out those who modified less than 10 times and labeled engineers as experienced or less experienced based on the 75th percentile of modifications. We randomly sampled and invited 30 engineers for interviews, of which 11 engineers agreed to participate (response rate of 36\%). Participants had modified on an average of 36 monitors over six months, with a minimum of 22 and a maximum of 75. We conducted 30-minute structured interviews with these 11 participants. We asked participants how often they modify their monitors and recorded their issues during the process. We introduced the proposed ontology and asked if it was helpful, and if they had any suggestions for new classes. We showed a sample recommendation and asked for a usefulness rating. Here are the questions and results from the study:

\myparagraph{Q1. How often do you create/update monitors?}
Here, a majority of seven participants said they edit monitors every month, while two others said they update them weekly. Interestingly, two other participants said they create or update monitors only on an yearly basis. 

\myparagraph{Q2. What challenges do you face while creating or updating monitors?}
From the user study, we observed that majority of the participant expressed problems in choosing and finding the right metrics and thresholds for the alerting logic. In addition, another major problem observed is finding and searching for similar duplicates. Here are some representative verbatim responses:
\textit{$P_{4}$ : "Creating the formula and finding the right metrics"}, \textit{$P_{5}$ : "Setting up threshold for alerts. It bothers me since I need to search historical data to find a good threshold. It should not be too loose or not to be too strict"}
\\
\myparagraph{Q3. Will you find a structured ontology for monitors, encapsulating the resource classes and SLO types, useful for creating or updating monitors?}
Here, notably, all participants said that they found the Ontology useful. One participant specifically described how the structure can be beneficial:\\
\textit{$P_{2}$ : "That will be very helpful. It'll be easy to search and organize monitors because over the life cycle of a product, there are too many monitors to manage and many older monitors go unnoticed."}

\myparagraph{Q4. Do you think AI based recommendations for creating/updating monitors would help? If no, why?}
Here, a majority of eight 
participants said that a recommendation system would be helpful. Only three participants felt some uncertainty that it maybe be useful.   

\myparagraph{Q5. How useful do you find this sample recommendation from our Intelligent monitoring framework?}
Here, we observed an average rating of 4.27 out 5 for the usefulness of the recommendations provided. Three participants responded that the monitor recommendations are extremely useful, while seven participants responded that it is somewhat useful. Only one person responded with neutral and no participant responded that it would not be useful.  

\section{Related Work}

\paragraph{Incident detection and management}
This area of research focuses on faster detection and mitigation of incidents \cite{li2021fighting,zhao2020automatically,bhattacharyya2016semantic,li2022intelligent}. Ref.~\cite{li2021fighting} discusses the issue of incidents and outages degrading the availability of large-scale cloud computing systems such as AWS, Azure, and GCP and proposes an automatic incident detection system for incident management, which alerts from different services and detects the occurrence of incidents from a global perspective. Similarly, \cite{zhao2020automatically} presents a framework for identifying severe alerts that extracts a set of interpretable features and identifies the severe alerts out of all incoming alerts. Ref.~\cite{li2022intelligent} summarizes the pain points and challenges in the real-world detection workflow for critical incidents, covering multi-aspect detection, proactive detection, and incident refinement. 
Once we have the framework to suggest appropriate monitors, these approaches can be integrated into our proposed framework to further improve incident detection and mitigation.
\paragraph{Service Level Objectives in Cloud Systems}
Literature on Service Level Objectives can be broadly classified into two categories. The first category defines the SLOs from a cloud perspective and provides an empirical study on the challenges associated with SLOs \cite{mogul2017thinking,mogul2019nines,jayathilaka2015response}. Ref.~\cite{mogul2019nines} discusses the concept of SLOs, SLA, and the challenges of defining Service Level Objectives (SLOs) in cloud systems. Cloud providers find it technically hard to meet the strong and understandable promises for reliable and adequate performance that cloud customers want, especially in the face of arbitrary customer behavior and hidden interactions. Refs.~\cite{ding2019characterizing,ding2020characterizing} address common misconceptions and highlight the practical aspects and challenges of defining and managing SLOs in the cloud. These works discuss practical considerations in defining and managing SLOs, such as defining SLOs at an appropriate level of granularity, understanding and managing service dependencies, and implementing effective monitoring and alerting systems to track SLO compliance and respond to deviations promptly.

The second category of literature discusses models, algorithms, runtime mechanisms, and tools for providing and consuming cloud resources in an SLO-native manner while offering performance guarantees to the user \cite{patros2017slo,nastic2020sloc,qiu2020firm,wang2022autothrottle}. While these works represent a frontier in defining the concepts of SLOs and their challenges in a cloud setting, they do not provide an extensive set of metrics or SLOs to be monitored by a cloud provider to ensure the continuous availability, efficiency, and reliability of a service. Most of the works assume that the specific set of SLOs to be monitored for a cloud service is already known.

\paragraph{Monitoring system for Cloud}
Refs.~\cite{surianarayanan2019cloud,montes2013gmone,aceto2013cloud} discuss the essentials of cloud monitoring in general and describe the characteristics of a good monitoring system. They also cover the various stages of monitoring, such as data collection, analysis, and decision-making, and list commercial and open-source tools available for monitoring, such as VMware Hyperic, Virtualization Manager of SolarWinds, and Azure Monitor. 
Ref.~\cite{nair2015learning} presents a hierarchical monitoring framework for Microsoft's internal distributed data storage and batch computing service. It discusses the application of machine learning algorithms for issue detection and highlights the advantages of machine learning in terms of adaptability and the ability to identify subtle anomalies in system behavior. Similarly, Ref.~\cite{klaver2021towards} proposes an early-stage monitoring solution called the Cloud Monitor, which aims to integrate existing and new benchmarks in a flexible and extensible way. These solutions, however, are application-oriented, focus on one or more applications, and develop some heuristics and/or data mining-based approaches to solve a selected set of problems. Similar solutions are developed for online services and infrastructure scenarios in \cite{fu2012performance,lin2014unveiling}.

The works discussed so far either develop approaches that work for specific applications or assume that ``what'' resource to monitor and ``which'' metric to monitor are known, and then develop techniques that can help in faster incident detection or alert refinement. The latter works are not directly related to our work, but once we solve the fundamental problem of monitor design and distribution, these approaches can be augmented to the proposed framework. 

\normalfont
\section{Discussion and Conclusion}
Continuous monitoring is crucial for maintaining the performance, reliability, and efficiency of cloud services. However, the current ad hoc monitoring practices in the industry can result in inefficiencies and inconsistencies. To address this issue, we have proposed a framework for recommending monitors 
based on their properties. We derived a structured ontology of the monitor space by mining data of over 30k monitors from 791 production services at \CompanyX{} and conducted an extensive empirical study. Our study provided valuable insights into the influence of service properties on the monitor ontology, which we used to propose a monitor recommendation framework. By leveraging the proposed framework, cloud service providers can ensure that their monitoring practices are consistent, efficient, and effective. We have evaluated the effectiveness and usefulness of the framework through user studies, which demonstrated that the framework can significantly improve the monitoring practices of cloud service providers.
Future studies will involve extending recommendations for existing services with additional signals, such as incident history, and refining the ``Service-level'' resource class into subclasses while identifying better features. Overall, our work provides a foundation for improving the monitoring practices of cloud services and enhancing their performance, reliability, and efficiency.



\end{document}